\documentclass[12pt]{article}

\usepackage{a4}

\begin{document}

\title{Static, Self-Gravitating Elastic Bodies}

\author{Robert Beig\\Institut f\"ur Theoretische Physik der Universit\"at
Wien\\ Boltzmanngasse 5, A-1090 Vienna, Austria\\[1cm] Bernd G. Schmidt\\
Max-Planck-Institut f\"ur Gravitationsphysik\\ Albert-Einstein-Institut\\
Am M\"uhlenberg 1, D-14476 Golm, Germany}
\maketitle

\begin{abstract}
There is proved an existence theorem, in the Newtonian theory,
for static, self--gravitating, isolated bodies composed of
elastic material. The theorem covers the case where these
bodies are small, but allows them to have arbitrary shape. 

Keyword: self-gravitating elastic bodies

\end{abstract}

\section{Introduction}
\label{introduction}

Most solutions of the Einstein field equations --- whether known
explicitly or given by existence theorems --- describing static
isolated bodies are spherically symmetric. The reason for this
is the choice of matter model which usually is that
of a perfect fluid - and such models are necessarily spherically
symmetric. This latter statement has been proved by
Lichtenstein \cite{lichtenstein}  in the Newtonian theory and  - in  the same
generality - is still a conjecture  in General Relativity.(For the
best results available, see \cite{beig}, \cite{lindblom}).
The only nonspherical solutions known to us are the axially symmetric ones
constructed by Rein for Vlasov Matter \cite{rein}. In the present work
we pursue another way to describe
nonspherical gravitational fields by choosing as our
matter model elastic bodies, coupled to the
static Einstein equations. In the spherically symmetric case this
has been done by Park \cite{park}. In the nonspherical case nothing is
known in the Einstein theory nor --- to the best of our
knowledge --- in the Newtonian theory.
Thus, as a first step, in the present paper we prove an existence theorem, 
purely in the Newtonian theory,
for static self-gravitating bodies
composed of elastic material. The theorem allows these bodies to have arbitrary shape.

The main limitation of the present work is that we restrict ourselves to
solutions close to the natural state of the body, which, in physical terms, means that
we require the body to be ``sufficiently small''. The main technical tool, then, is
the implicit function theorem near that natural state. For ``pure traction problems''
such as the one studied here, there often occur phenonema of non-uniqueness beyond
the trivial one stemming from invariance under Euclidean motions. These phenomena,
which have been thoroughly studied (see Chillingworth et al \cite{chill})
do not happen in the problem at hand. The reason is that  traction problems 
require, e.g. in the case of vanishing
traction, certain compatibility conditions (``equilibration conditions'') on the load, namely that the total
force and total torque it exerts on the body be zero. In our case, where the load
is given by the pull of the body's own gravitational field, these quantities are
a priori zero.

 We also point out that we are not able to make statements on the global problem, 
i.e. what happens
far away from the natural state. For this one would invoke variational techniques, in
particular the powerful methods introduced into the subject by Ball (see e.g.
\cite{ball}).

Let $\bf{B}$ be an open, bounded, connected subset of
$\bf{R^3}$ with smooth boundary. The domain $\bf{B}$ (``body'') is our reference configuration. 
We also consider 1-1 maps $\phi: \bf{B}\rightarrow\bf{R^3}$ (``physical space''),
$x_i=\phi_i(X_A)$. Let $\bf{B^{\phi}}\subset \bf{R^3}$ be the image of $\bf{B}$ under $\phi$.
Then the basic field equations are as follows:

\begin{equation}\label{div} 
-div_x T^{\phi} = \rho\; grad_x U  \hspace{1.5cm}   \mbox{in}\;\bf{B^{\phi}}     
\end{equation}
\begin{equation}\label{lap}
\Delta_x U = 4 \pi G \rho	\hspace{2cm}	\mbox{in}\; \bf{R^3}
\end{equation}
Here $G$ is the Newton
constant, $U$ is the gravitational potential, $T^{\phi}$ is the symmetric Cauchy
stress tensor, the mass density $\rho$ satisfies $\rho = n \rho_0$ with $\rho_0$ a 
positive constant 
in $\bf{B^{\phi}}$ and $\rho_0 \equiv 0$ in $\bf{R^3}\setminus \bf{B^{\phi}}$. Finally,
the number density $n: \bf{B^{\phi}}\rightarrow \bf{R^3}$ is given by $n(x)= \det \nabla
f(x)$, where $f$ is the inverse map of	$\phi$. We assume $n > 0$ in
 $\overline{\bf{B^{\phi}}}$. Let us remark that Equ.(\ref{div}) and Equ.(\ref{lap})
 also describe perfect fluids,
namely if it is assumed that $T^{\phi} = pI$, where the pressure $p$ is a function
just of $n$ and $I_{ij} = \delta_{ij}$.

We first make some observations on the equations (\ref{div}) and (\ref{lap})
separately. The divergence structure of (\ref{div}) implies that its right-hand
side, say $b$, satisfies a compatibility condition, as follows: 
Let $\xi_i (x)$ be a Killing vector of $\bf{R^3}$, considered as flat Euclidean
space, i.e. of the form

\begin{equation}\label{killing}
\xi_i (x) = c_i + \omega_{ij}x_j \Leftrightarrow \partial_i \xi_j + \partial_j\xi_i = 0, 
\end{equation}
where $c_i$ and $\omega_{ij} = \omega_{[ij]}$ are constants. Then, upon scalar
multiplication of (\ref{div}) with $\xi$ and integrating over ${\bf{B^{\phi}}}$,
 we easily find that

\begin{equation}\label{compatibility}
\int_{\bf{B^{\phi}}} b \cdot \xi + \int_{\partial\bf{B^{\phi}}}
t^{\phi} \cdot \xi = 0,
\end{equation}
where $t_i^{\phi} = T_{ij}^{\phi}\nu_j^{\phi}$
 with $\nu^{\phi}$ the outward unit-normal of
$\partial\bf{B^{\phi}}$.

On the other hand the ``load'' $b$ inserted on the r.h.side of (\ref{div})
has the property that it gives no contribution to Equ.(\ref{compatibility}).
This is seen as follows. Define a symmetric 2-tensor $\Theta$ by 

\begin{equation}\label{theta}
\Theta = \frac{1}{4\pi	G} ( grad_x U \otimes  grad_x U -
 \frac{1}{2} I\; grad_x U \cdot grad_x U ).
\end{equation}
Then Equ. (\ref{lap}) implies 

\begin{equation}\label{div1}
div_x \Theta = \rho \; grad_x U 
\end{equation}
Suppose, in addition, that $U$ satisfies

\begin{equation}\label{order}
U = O(\frac{1}{|x|}) \hspace{2.5cm} \mbox{on}\; \bf{R^3}\setminus\bf{B^{\phi}}
\end{equation}
Operating with $\xi$ on (\ref{div1}) as before on (\ref{div}), but with integration
 over $\bf{R^3}$, we find that the load $b$
used in Equ.(\ref{div}) is ``automatically equilibrated'' in the above sense.
Put differently, choosing for $\xi$ the three translation Killing vectors, this
statement amounts to saying that the force exerted on the body by its own
gravitational field is zero. Similarly, using three rotational Killing vectors,
implies the vanishing of the gravitational self-torque. 

We want to solve the coupled system (\ref{div}) and (\ref{lap}) subject to 
no-traction boundary conditions,
namely that $t_i^{\phi}$ be zero on $\partial\bf{B^{\phi}}$, which is a free boundary.
To make the problem tractable it is thus important to write the above equations
as PDE's on  $\bf{B}$, rather than $\bf{B^{\phi}}$, using the Piola transform.
With the definition $T_{iA} = n^{-1} f_{A,j} T_{ij}^{\phi}$, one finds (see e.g.
\cite{ciarlet}) that

\begin{equation}\label{div3}
 - div_X T = \rho_0\; grad_x U.
\end{equation}
If $T_{ij}^{\phi}$ is solely a function of $f_{A,i}$, $T_{iA}$ can be viewed as a function
of $(\nabla\phi)_{i,A} = F_{i,A}$. This follows from the chain rule for
differentiation. The potential $U(x)$, satisfying (\ref{lap}) and (\ref{order}), is
given in physical space by

\begin{equation}\label{newton}
U(x) = -\;G\rho_0\;\int_{\bf{B^{\phi}}} \frac{n(x')}{|x-x'|}\;d^3x'.
\end{equation}
Consequently (\ref{div3}) takes the form

\begin{equation}\label{div4}
- \partial_A T_{iA} = G\rho_0 \int_{\bf{B}} \frac{\phi_i(X) - \phi_i(X')}
{|\phi(X) - \phi(X')|^3} d^3X'
\end{equation}
and the compatibility conditions (\ref{compatibility}), using (\ref{killing}), result in

\begin{equation}\label{c1}
\int_{\bf{\partial B}}t = 0
\end{equation}
\begin{equation}\label{c2}
\int_{\partial\bf{B}} t \wedge \phi (x) = 0,
\end{equation}
where $t_i = T_{iA} \nu_A$. Our aim is to solve (\ref{div4}) for $\phi$, subject to the
boundary conditions

\begin{equation}\label{b}
t| _{\partial\bf{B}} = 0.
\end{equation}
We assume that  

\begin{itemize}
\item[A1]
$T_{iA}(\nabla \phi) = 0$, when $\nabla \phi = I$
\item[A2]
The linearization-at-($\phi = id$) of the operator $div_XT$ is strongly elliptic, in
other words, $a_{iAjB} = \frac{\partial T_{iA}(F) } {\partial F_{jB}}$ satisfies
$a_{iAjB} = a_{jBiA}$ and $a_{iAjB}|_{F=I} v_i v_j V_A V_B \geq0$
\end{itemize} 
   
The physical meaning of condition (A2) is as follows: The natural state is
usually supposed to be such that $a_{iAjB}|_{F=I} = \mu (\delta_{ij}\delta_{AB} +
\delta_{Aj}\delta_{iB}) + \lambda \delta_{iA}\delta_{jB}$ for constants $\mu$
and $\lambda$, the Lam\'{e} moduli. The ellipticity condition (A2) is then
equivalent to the inequalities $\mu > 0, 2\mu + \lambda >0$. A different 
interpretation of (A2) would be by saying that plane waves propagating according
to the linearized-at-$F = I$ time dependent equations have real frequency.
  
We note that (at least) condition (A2) rules out fluids. And, indeed, the theorem
of the next section stating the existence of bodies of arbitrary shape, can not
possibly apply to perfect fluids, as noted in the Introduction. 

 \section{The Main Theorem}

\label{main theorem}

 We now state our precise assumptions. As configuration space $\bf{{\cal C}}$
we take maps $\phi:\bf{B}\rightarrow \bf{R^3}$ with $\phi_i \in W^{2,p}({\bf B})^3,\, p>3$,
and in it $\mathbf{{\cal C}}_{\epsilon} \subset \mathbf{{\cal C}}$ of maps
 $\phi_i = X_i + h_i$ with $\parallel h \parallel_{2,p} < \epsilon$. For $\epsilon$
sufficiently small, $\phi$ is $C^1$ - map close to the identity with $C^1$ - inverse
(see Appendix). For the stress tensor $T_{iA}(\nabla \phi)$ we assume that it is
in $C^2({\bf{R^9}},{\bf{R^9}})$ and that it satifies conditions $(A1,2)$ of
section  \ref{introduction}, wherefrom it follows \cite{valent} that the operator
$\phi \mapsto T(\nabla \phi)$ is a $C^1$ - mapping from $W^{2,p}({\bf{B}})^3$ to
$W^{1,p}({\bf{B}})^9$. Our main result is
 
{\bf Theorem}: For sufficiently small $G$ there is a solution $\phi \in {\cal C}_{\epsilon}$
 of (\ref{div4}) subject to (\ref{b}). This solution is unique provided

\begin{equation}\label{unique}
h_i(\vec{0}) = 0,\;\partial_{[i}h_{j]}(\vec{0}) = 0.
\end{equation} 
(We assume that $\vec{0} \in \bf{B}$.)

We remark that the smallness-condition for $G$ can of course, by scaling, be
rephrased by $G \rho_0 \ll \frac{|T|}{L^2}$, where $L$ is a typical length
scale of $\bf{B}$ and $|T|$ an upper bound for the stress tensor.
 Our method of proof follows the geometrical
treatment of the Stoppelli theorem \cite{stoppelli} due to LeDret \cite{ledret}.

{\bf{Proof}}: Consider the map $E:{\cal C} \mapsto {\cal Y} = \{(b,t)
 \in W^p({\bf{B}})^3 \times W^{1-1/p,p}(\partial{\bf{B}})^3, p>3\}$,
 the operator of nonlinear
 elasticity given by

\begin{equation}\label{elasticity}
 \phi \mapsto (-div_X T(\nabla \phi), T(\nabla \phi) \nu)
\end{equation}
 The operator $E$ is well-defined and $C^1$ (see \cite{ledret}). Now recall from the
discussion of Section \ref{introduction} 
that elements $(b,t) \in {\cal Y}$ lying in $E(\phi)$ satisfy the compatibility 
(``equilibration'') conditions, namely

\begin{equation}\label{e1}
\int_{\bf{B}} b + \int_{\partial\bf{B}} t = 0
\end{equation}
\begin{equation}\label{e2}
\int_{\bf{B}} b \wedge \phi(X) + \int_{\partial\bf{B}} t \wedge \phi(X) = 0
\end{equation}

The set ${\cal Y}_{\phi}$ of pairs $(b,t) \in {\cal Y}$ satisfying (\ref{e1}) and (\ref{e2}),
for given $\phi$, is a vector subspace of ${\cal Y}$ of codimension 6, when
 $\phi \in {\cal C}_{\epsilon}$. (More precisely, it follows from the results of LeDret,
and is easy to check, that the only elements $\phi \in {\cal C}$, at which
 ${\cal Y}_{\phi}$ fails to have codimension 6, are those for which the image $\phi(\bf{B})$
 is parallel to a fixed direction $v \in \bf{R^3}$ - which is impossible if 
$\phi \in {\cal C}_{\epsilon}$.) Let us choose some complement $S$ of
 $L_e \subset {\cal Y}_e$, where $L_e = {\cal Y}_{id}$ and define a projection
$P: {\cal Y}_{\phi} \mapsto L_e$. The linear maps $P: {\cal Y}_{\phi} \mapsto L_e$
are isomorphisms and $C^1$ (see \cite{ledret}, proof of Proposition 1.4). Next
consider the (``live'') load afforded by the gravitational 
force, i.e.  $b = G \bar{U}(\phi)$ with $G \bar{U}_i(\phi)$ given by the right-hand side
of (\ref{div4}). By explicit calculation, or from the discussion of section\ref{introduction},
it follows that $\bar{U}(\phi) \subset {\cal Y}_{\phi}$. Note that this requires
$\bf{B}$ to be connected. If $\bf{B}$ had several connected components,
 $\bar{U}(\phi)$ would be automatically equilibrated only with respect to the
whole of $\bf{B}$, whereas (\ref{e1}) and (\ref{e2}) would for the operator $E$
be required to hold separately for each connected component of $\bf{B}$. It is
thus important that we have only one body.

We want to solve the equation

\begin{equation}\label{equation}
E(\phi) = G \bar{U}(\phi) \;\mbox{on}\; {\bf{B}}, \hspace{0.7cm} t_i = 0 \;\mbox{on}\;
 \partial\bf{B}
\end{equation}
for small $G$. We know from A1 that $\phi = id$ is a solution for $G=0$.
We write

\begin{equation}\label{operator}
F(G,\phi) = E(\phi) - G\bar{U}(\phi),
\end{equation}
with $F$ viewed as a function ${\bf{R}} \times {\cal C}_{\epsilon} \mapsto {\cal Y}$.
In the Appendix we show that $\bar{U}$, whence $F$, maps ${\cal C}_{\epsilon}$ into
${\cal Y}$ in a $C^1$ - fashion. If we now compute the linearization of $F$ at 
 $\phi = id$ for $G=0$, we find that this is a map from
${\cal C}$ to ${\cal Y}$ which is not surjective, due to the presence of the 
equilibration conditions. To get round this difficulty, consider the modified operator

\begin{equation}\label{modified} 
F'(G,\phi) = P(E(\phi)) - G P(\bar{U}(\phi),
\end{equation}
with $F'$ viewed as a map ${\bf{R}} \times {\cal C}_{\epsilon} \mapsto L_e$. Clearly
every solution of $F'=0$ is also a solution of $F=0$. If, in addition, we eliminate
the translational and rotational freedom by replacing ${\cal C}$ by
 ${\cal C}_{sym}$, consisting of all elements $\phi_i = X_i + h_i$ in ${\cal C}$
for which $u_i(\vec{0}) = 0$ and $\partial_{[i} u_{j]}(\vec{0}) = 0$, it follows
from standard linear theory (see \cite{marsden}, Lemma 3.17 of Chap.7), that the 
linearization-at-($\phi = id)$ of $F'$ at $G = 0$ is an isomorphism
 ${\cal C}_{sym} \mapsto L_e$. Hence our claim follows from the implicit function
theorem. 

\section{Appendix}
\label{appendix}

The Poisson integral used in the body of the paper is given by

\begin{equation}\label{poisson} 
(\bar U_i[\phi])(X)=\int_{\bf{B}}\frac{\phi_i(X)-\phi_i(X')}
{|\phi(X)-\phi(X')|^3} d^{3}X'
\end{equation}
Here ${\bf{B}}\subset {\bf{R^3}}$ is bounded, not empty, open and connected with  $\partial\bf{B}$
 smooth and $\phi_i\in W^{2,p}({\bf{B}})^3$, $p>3$. Furthermore
$\phi_i(X)= X_i+h_i(X)$, and we assume that $\parallel
h\parallel_{2,p}<\epsilon$, $\epsilon$ small. It follows from Sobolev
embedding that $\partial_jh_i$ is small , in particular bounded in $\bf{\bar{B}}$.
By the mean--value theorem we infer that

\begin{equation}\label{h-h}
|h(X) -h(X')|< C' |X-X'|
\end{equation}
This is immediate for $\bf{B}$ convex, otherwise see \cite{ciarlet}, p.
224.
Making $\epsilon$ smaller, if necessary, we have that

\begin{equation}\label{delh}
|\partial h(X)|<{1\over 2}
\end{equation}
Consequently, there exist positive constants $E,E'$ such that 

\begin{equation}\label{e}
E|X-X'|\leq |\phi(X)-\phi(X')|\leq E'|X-X'|
\end{equation}
It immediately follows that $\bar U_i[\phi]\in C^0({\bf{\bar{B}}})^3$.
Thus $\bar U_i$ is a bounded map of ${\cal U}_\epsilon(id_{\bf{B}})\subset
W^{2,p}({\bf{B}})^3\to
C^0({\bf{\bar B}})^3$, whence to
to $W^{0,p}({\bf{B}})^3$. Here ${\cal U}_\epsilon(id_{\bf{B}})$ denotes the set of $\phi$'s in
the $\epsilon$--ball centered at the identity map. 
 We want to show $\bar{U}_i$ is actually $C^1$.
We first compute the Gateaux--derivative (directional derivative) of $\bar
U_i$, namely

\begin{equation}\label{gateaux}
\left[{d\over dt}\bar U_i[\phi^t]\right]_{t=0}= D\bar U_i[\phi^0]\cdot v
\end{equation}
where $ \phi^t_i(X)= X_i+tv_i(X), v_i\in W^{2,p}({\bf{B}})^3$.
First observe that the expression 

\begin{equation}\label{c}
{1\over t}\left\{
{\phi_i^t(X)-\phi^t_i(X')\over|\phi^t(X)-\phi^t(X')|^3}-
{\phi_i^0(X)-\phi^0_i(X')\over|\phi^0(X)-\phi^0(X')|^3}
\right\}
\end{equation}
for $X\neq X'$, converges pointwise for $t\to 0$ to 

\begin{equation}\label{con}
{v_i(X)-v_i(X')\over|\phi^0(X)-\phi^0(X')|^3}-
{3(\phi^0_i(X)-\phi^0_i(X'))(\phi^0_j(X)-\phi^0_i(X'))(v_j(X)-v_j(X'))
\over|\phi^0(X)-\phi^0(X')|^5}
\end{equation}
Next note the following chain of elementary inequalities: $b_1,b_0$
vectors
$\in \bf{R}^3$ (or $\bf{R^n})$

\begin{equation}\label{1}
\mid{1\over |b_1|}-{1\over |b_0|}\mid\leq{|b_1-b_0|\over|b_1||b_0|} 
\end{equation}
\begin{equation}\label{2}
\mid{1\over |b_1|^2}-{1\over |b_0|^2}\mid\leq\left({1\over |b_1|^2|b_0|}
+{1\over |b_0|^2|b_1|}\right)|b_1-b_0|
\end{equation}
\begin{equation}\label{3}
\mid{1\over |b_1|^3}-{1\over |b_0|^3}\mid\leq\left(
{1\over |b_1|^3|b_0|}+{1\over |b_1|^2|b_0|^2}+{1\over |b_1||b_0|^3}\right)
|b_1-b_0|
\end{equation}
$a_1,b_1,a_0,a_1$ vectors $\in \bf{R^n}$

\begin{equation}\label{4}
\left|{a_1\over |b_1|^3}-{a_0\over |b_0|^3}\right|\leq
\left(
{1\over|b_1|^3|b_0|}+{1\over |b_1|^2|b_0|^2}+{1\over |b_1||b_0|^3}
\right)
|a_1||b_1-b_0|
\end{equation}
\begin{equation}\label{5}
+{|a_1-a_0|\over|b_0|^3}
\end{equation}
setting 

\begin{equation}\label{6}
a_1=b_1=\phi^t(X)-\phi^t(X')
\end{equation}
\begin{equation}\label{7}
a_0=b_0=\phi^0(X)-\phi^0(X')
\end{equation}
we find

\begin{eqnarray}
\label{ssds}
\nonumber
\lefteqn{
\left| {\phi_i^t(X)-\phi^t_i(X')\over|\phi^t(X)-\phi^t(X')|^3}-
{\phi_i^0(X)-\phi^0_i(X')\over|\phi^0(X)-\phi^0(X')|^3}
\right|\leq } \\
& & \left(
{2\over|\phi^0(X)-\phi^0(X')|^3}
+{1\over|\phi^t(X)-\phi^t(X')|^2|\phi^0(X')-\phi^0(X)|}\right. \\ \nonumber
& &\left. +{1\over |\phi^t(X')-\phi^t(X)||\phi^0(X')-\phi^0(X)|^2}\right)t\
  |v(X')-v(X)|
\end{eqnarray}
It follows, using (\ref{e}), that the sequence in (\ref{c}) is bounded by a positive,
$t$--independent   function, whose integral over $X'\in \bf{B}$ is a bounded
function of $X$ in
$\bf{\bar{B}}$. So, by dominated convergence, the previous limit, whence the
Gateaux derivative actually exists. But the linear operator defined by the
directional derivative $v\in W^{2,p}({\bf{B}})^3\to W^{0,p}({\bf{B}})^3$, is clearly
bounded. So, by a standard theorem (see e.g. \cite{abraham}, Corollary 2.4.10),
${\bar{U}}(\phi)$ is a $C^1$--functional.

\vspace{0.5cm}

{\bf{Acknowledgments}}: The authors thank Horst Beyer and Alan Rendall for useful discussions.

\end{document}